# EQUITY-BASED INCENTIVES, PRODUCTION/SERVICE FUNCTIONS AND GAME THEORY

**Michael C. Nwogugu**
E-mail addresses: mcn2225@aol.com.
P. O. Box 996, Newark, New Jersey 07101, USA

**Abstract**
EBIs/ESOs substantially change the traditional production/service function, because ESOs/EBIs have different psychological impacts (motivation, or de-motivation), can create intangible capital (ie. Social Capital, Reputational Capital and Human Capital), and create different economic payoffs. Although Game Theory is a flawed concept, it can be helpful in describing interactions in ESO/EBIs transactions. ESOs/EBIs involve a two-stage game; and there are no perfect Nash Equilibria for the two sub-games. The large number of actual and potential participants in these games significantly complicates resolution of equilibria and increases the dynamism of the game(s), given that players are more sensitive to each other's moves in such games. This article: **a**) analyzes how ESOs/EBIs affect traditional assumptions of production functions (in both the manufacturing and service sectors), **b**) analyzes ESO/EBIs transactions using game theory concepts, **c**) illustrates some of the limitations of game theory.

1. ESOs/EBIs, Incentive Problems And Principal-Agent Theories.

In their present form, ESOs/EBIs create moral hazard and incentive problems by providing the wrong combinations of incentives, timing, managerial power, and opportunities to act – which result in earnings manipulation, inefficient capital budgeting decisions, excessive risk-taking, un-warranted wealth transfers, etc.. Nwogugu (2004). However, standard Principal-Agent theory is not always applicable in the ESO/EBI context because of the following factors. The optimal incentive contract is one that minimizes the costs of compliance, minimizes the propensity to commit fraud, maximizes the after-tax cash benefits to the grantee, maximizes shareholder value before compensating grantees and most accurately matches performance and reward. Many forms of EBIs/ESOs such as indexed EBIs/ESOs are not efficient. Incentive contracts can reduce moral hazard, agency and compensation-accuracy problems in companies.

The board of directors is typically influenced by management. ESOs/EBIs give management 'shadow' ownership interests in the company, which makes them behave more like shareholders. Managements' actions are increasingly being limited and controlled by boards of directors, employees, shareholders, activist groups, and some laws (such as the Sarbanes-Oxley Act in the US). The contracts between managers and shareholders are typically incomplete contracts, and the principals' (shareholders') property interests are incomplete, and thus principal-agency theory is inapplicable.

The nature of decision-making in the firm and the structure of the firm may change the principal-agent relationship. Many key decisions (manager compensation plans, reorganizations, acquisitions, raising new capital, etc.) are increasingly being shifted to shareholders and there is increasing shareholder activivism worldwide.

The legal/statutory and company-specific restrictions on equity-based compensation reduces its efficiency as a corporate governance and incentive tool. The lack of employee portfolio diversification implicit in ESO/EBIs awards creates substantial incentives for options grantee to commit fraud, and changes the principal-agent relationship. Employees' use of derivatives to hedge and monetize their equity compensation effectively renders principal-agent theories inapplicable, by changing the risk profiles, tax consequences, economic benefits and incentive effects of ESOs/EBIs. Most principal-agent models erroneously assumes that all types of risk are costly to managers and that such risk have the same value to both the principal and the agent. EBIs/ESOs typically have different values to managers, the company and shareholders/investors.

Psychological and non-monetary motives can affect the structuring of, and the employee reaction to incentive compensation and ESOs/EBIs, and thus render principal-agent theory completely inapplicable. Such motives include: 1) the desire to gain specialized experience or work on interesting projects, 2) desire to avoid social rejection, 3) reciprocity, 4) commitment to public service, 5) personal experiences, 6) employee's socio-economic background may limit his/her ability to negotiate for adequate compensation, 7) fear of uncertainty which may cause employees to prefer combinations of certain and uncertain compensation. The employer may be influenced by other motives such as: 1) job preservation, 2) cross-training employees for assignment to other divisions, 3) labor union concessions, 4) prevention of unionization.

2. Existing Literature.

This article doesn't advocate game theory as an accurate representation of real-world or theoretical interactions or bargaining. Indeed, several authors have found game theory to be inaccurate – see: Brams (2000); Bromiley & Papenhausen (2003); Chatterjee (1996); Goeree & Holt (2001); Green (2004); Hechter (1992); Stone (2001); Troger (2005); Conitzer & Sandholm (2008); Carmona (2004); Kalai (2004); Camerer (2003); Denzau & Roy (March 2005).

Financial Markets are now significantly maintained/driven and influenced by computer systems and are Large-scale systems. The relevant literature on incentives and Game Theory in Large-Scale Systems include the following: Katz (1990); Yoon (2004); De Clippel (2005); Courty & Marschke (2004); Nwogugu (2004); Lipman & Wang (2000); Sung (2005); Kaplan & Henderson (2005); Birk (1999); Allen (2003); Milnor & Shapley (1978); Aumann (1975); Butnariu (1987); Hart & Mas-Colell (1995); Owen (1972); Moulton (2000); Wrobleski (2007); Mehran & Rosenberg (October 2007); Black & Lynch (2004); Calomiris & Hubbard (Jan. 2004); Gandolfo (2008); Pickard, Kitchenham & Jones (1999); Lee & Keen (2004); Baier, Dwyer & Tamura (2002); Ackerberg, Caves & Frazer (Nov. 2005); Fehr, Fischbacher & Kosfield (2005); Pekkarinen (2002); Growiec (2008); Fioretti (2007); Neill (2003); Santin (2008); Wall (1998); Kuhl (2005); Jones & Kato (1995); Wadhwani & Wall (1990); Thoron (2004); Tseo, Sheng, Peng-Zhu & Lihal (2004); Mariotti (1997); Ehtamo & Hamalainen (1989); Ehtamo, Kitti & Hamalainen (2002); Groves & Loeb (1979); Heiskanen, Ehtamo & Hamalainen (2001); Zheng & Balsar (1982); d'Aspremont & Gérard-Varet (2002); Guesnerie & Laffont (1984); Myerson (1979); Maskin & Tirole (1992); Holmstrom & Milgrom (1987); Holmstrom & Milgrom (1991); Hirshleifer & Rasmusen (1992); Miller & Whitford (2002); Armstrong (2002).

The omissions and problems in existing game-theoretic analysis of ESOs/EBIs include the following: a) the implicit assumptions of game theory are not realistic and often cannot be expressed in quantitative real-life terms, b) the models/theories erroneously assume constant alignments of economic interests and incentives between management and employees, between the board of directors and management, between advisors and shareholders, and between venture-capital companies and shareholders; c) the models/theories erroneously assume that there are no or minimal monitoring costs, transaction costs and compliance costs, d) most game-theoretic analysis omits the utility/disutility that an employee gains from owning EBIs/ESOs, and the utility/disutility thata company gains from issuing EBIs/ESOs.

ESOs/EBIs create two-stage multi-person games that are non–zero-sum games. The two stages of the 'game' are characterized as follows.

3. ESOs/EBIs and Game Theory.

This section introduces some new theories about ESOs/EBIs and game theory; and attempts to explain EBI/ESO dynamics within existing game theory concepts.

Stage-One.
The first stage of the game occurs during the process of negotiating the terms of issuance of ESOs/EBIs; and it involves the employee, the company, management, venture capital firm (or major seed capital investor), third-party compensation advisor (only in the case of established companies) and the board of directors. The first stage begins immediately after the company's Board of Directors determines and approves the general terms of EBIs and the associated incentive plan – in some jurisdictions, the shareholders may also have to approve the terms of incentive plans. The first stage ends when the employee and the firm signs the EBI contract. Typically, most employees

(except executives) don't have any influence on the pre-game negotiations and contracting pertaining to the general terms of the compensation program. Hence, manager with such knowledge will have an information advantage over many employees.

This stage-one game is repeated many times for different employees. During the first stage, the employer makes an offer to the employee, and the employee can either accept the EBI contract, continue negotiations, quit the job or look for another job. The employee's utility function may or may not be known to the firm and to other players. Similarly, the other players may or may not know firm's utility function.

The game is a Multi-person, non-zero-sum game. The rules of the game are set by participants, but players actions and expectations are constrained by various rules such as the entity's corporate bylaws or operating agreement, company policies/procedures, disclosure statutes, accounting standards (FASB;IASB); state/federal corporation law statutes, state/federal securities laws, and tax laws. The game is played only once (just before, and at issuance of EBIs/ESOs). Transaction costs are relatively minimal but monitoring costs are substantial.

Other employees may learn about the terms of the EBIs that are issued to one employee or to a class of employees. Upon obtaining such information, their reaction and strategies will depend on whether their employment is at-will, by a single contract or is covered by a collective bargaining agreement (which typically don't affect EBIs). Such informed employees may choose to leave the firm, form coalitions with other employees to exchange information and or negotiate for more EBI compensation, or just accept the news.

There is some measure of irreversibility because most EBI agreements are standard – repricing or back-dating of EBIs is now illegal in most jurisdictions.

There can be Nash Equilibria. The 'apparent' Nash Equilibria is the EBI/ESO Agreement (between the company and the employee). However, players may diverge from this 'apparent' Nash Equilibria due to the incentive effects of EBIs/ESOs and the uncertainty of valuation methods (for valuing EBIs/ESOs). <u>The real Equilibria is a mildly fluctuating state that is not continuous over time, and is determined by employee performance, company performance, vesting terms, communication among employees, and the nature of the EBI/ESO agreement.</u> The game is typically a cooperative game, in which one party (the company) has substantial negotiating leverage. Thus, Coalitions are possible, and typically occur.

There is no real prisoner's dilemma because: a) the shareholders, management and the board of directors typically approve the general terms of the employee incentive plan, b) the negotiations for the grant of ESOs/EBIs to any employee typically last for more than one day, c) employees typically have opportunities to exchange information about EBIs/ESOs and specific incentive compensation packages.

The value of this game varies for each player and consists of: a) the value of the EBI/ESO, b) transaction costs, c) monitoring costs, d) compliance costs.

Let:
$V_c$ = the value of EBI/ESO to the Company.
$V_e$ = the value of EBI/ESO to the employee. $V_e$ differs for each employee even where each employee holds the same quantity of EBIs – this is because of differences among employees pertaining to risk aversion, motivation, incentive effects, beliefs about the company's prospects, knowledge, perceptions of disclosed information, etc..
$T_c$ = the company's transaction costs – legal fees, staff costs, overhead/administrative costs, etc..
$T_e$ = the Negotiating-Employee's transaction costs – legal fees, employee time, administrative costs, etc.. $T_e$ varies for each employee even where each employee holds the same quantity of EBIs – this is because of differences among employees pertaining to choice of brokerage firms, knowledge, risk aversion, beliefs about the company's prospects, perceptions of disclosed information, etc..
$M_c$ = the company's monitoring costs.
$M_e$ = the Negotiation-Employee's monitoring costs.
$C_c$ = the company's regulatory compliance costs.
$C_e$ = the Negotiating-Employee's regulatory compliance costs.
$U_e$ = the Negotiating-Employee's utility from holding EBIs/ESOs. $U_e$ is different from $V_e$, and neither is a sub-set of the other. $U_e$ varies for each employee even where each employee holds the same quantity of EBIs – this is because of differences among employees pertaining to risk aversion, motivation, knowledge, incentive effects, beliefs about the company's prospects, perceptions of disclosed information, etc..

$U_c$ = the company's utility from awarding EBIs/ESOs – which includes increased employee morale/motivation, increased employee productivity, expectations, increased employee commitment, Social Capital, reduced stress among managers, risk management, etc.. $U_c$ is different from $V_c$, and neither is a sub-set of the other.
$\gamma$ = number of employees in the compensation program.
$H$ = time horizon during which game is played.
$L_c$ = present value of tax related disadvantages/losses incurred by company due to issuance and or exercise of EBIs/ESOs.
$L_e$ = present value of the tax related disadvantages/losses incurred by the employee due to issuance and or exercise of EBIs/ESOs.
$\mathcal{E}_a$ = the employee's actual effort level – measured in terms of productivity, quality of work and conformity.
$\mathcal{E}_r$ = the employee's minimum required effort – measured in terms of productivity, quality of work and conformity.
$\Lambda_e$ = estimated present value of proportionate equity dilution-related losses incurred by the employee when other employees exercise their EBIs.
$\Lambda_c$ = estimated present value of the equity dilution-related losses incurred by the company when the employee exercises his/her EBIs.
$\Omega$ = a vector that relates the number of employees that own EBIs/ESOs, to the value of the game to the firm; such that the term $\{(U_c + \mathcal{E}_a + V_c - T_c - \Lambda_c - M_c - L_c - C_c)^{(\pi*\Omega)}\}$ is a proxy for the employee's proportional share of the value of the game to the firm.

$\pi$ = a vector that accounts for: 1) the marginal rate of substitution of 'original' labor with capital (automation, technology, inputs, etc.), 2) the marginal rate of substitution of 'original' labor with new domestic labor – relevant where new employees can be trained quickly, 3) the marginal rate of substitution of 'original' labor with foreign labor, that require much less incentives, 4) the marginal propensity to substitute labor with automation; and 5) the firm's financial capacity to actually substitute 'original' labor. The marginal rate of substitution of labor is inversely proportional to the value of the game to the firm. $\pi$ has at least two values which depend on whether the objective is to calculate the values of the games to the employee or to the firm.

$\psi$ = a vector that accounts for the employee's deviation from the average or minimum required effort. The employee's actual effort level may not be observable by the firm. Furthermore, managers may have incentives to over-state or under-state employees' reported effort levels. $\psi$ has at least two values which depend on whether the objective is to calculate the values of the games to the employee or to the firm.

$\lambda$ = a vector that accounts for: a) the probability of monetization of the EBI by the employee (using derivatives and other methods), b) the probability that the employee will hedge the EBI, so that the EBI will not be sensitive to declining share prices. $\lambda$ is directly proportional to the value of the game to the company. $\lambda$ has at least two values which depend on whether the objective is to calculate the values of the games to the employee or to the firm. The game players may have Transferable Utilities ("TU") or non-transferable utilities ("NTU") depending on: a) the employee's knowledge and resources and ability to monetize EBIs before or after vesting, b) whether the EBI contract permits partial or full transfer of the EBIs, c) whether the firm can transfer or assign the EBIs if a major corporate transaction occurs, d) switching costs – costs of finding a new job or recruiting a new employee, costs of re-negotiating EBI contracts.

The solution to this N-payer game is a set of imputations that cover maximization or minimization of option value, transaction costs, and monitoring costs. The value of game to the company is:

$G_c = [U_c + \mathcal{E}_r - V_c - C_c - \Lambda_c - M_c - L_c - T_c]^{(\pi*\lambda)}$;

The Company's multi-objective function will be:
Min $\int_0^{Tc} \{V_c + T_c + M_c + L_c + C_c\}^{(\pi)} \partial T_c$.
Min $(\mathcal{E}_r - \mathcal{E}_a)$
Max$[U_c + \mathcal{E}_r - V_c - C_c - \Lambda_c - M_c - L_c - T_c]$

The value of the game to employee will be:

$$G_e = \text{Max} [\text{Max}(\{_0\int^{Tc}\{(U_c + \epsilon_a + V_c - T_c - \Lambda_c - M_c - L_c - C_c)^{(\pi*\Omega)}\}\partial T_c \},0), \{_0\int^{Te} \{U_e + V_e - \epsilon_a - T_e - \Lambda_e - M_e - L_e - C_e\}^{(\pi*\lambda)} \partial T_e\}]$$

    The following are the characteristics of this game, some of which contravene established Game Theory principles.  Any equilibrium solution (including the Nash Equilibria solution, if any), partly depends on the utility function of the players and occurs only over small intervals of time (not continuous).  Two or more players can improve any equilibria solution (or an assumed Nash Equilibria solution) simultaneously by varying their efforts, by increasing communication, and by disclosing their negotiation strategies and post-negotiation strategies to other players – post-negotiation strategies include varying their effort levels, monetization of EBIs, hedging EBIs, approving riskier projects, cost-cutting, share-repurchases, selective disclosure of information, approval of accounting accruals and changes, etc..  Any equilibria solution is not symmetric - because reversing the roles of the players will affect the solution.  Any equilibrium solution will be affected by the introduction/negotiation of irrelevant alternatives – such occurrences change players' motivations, incentive levels, strategies, efforts and ultimate objectives.  Not all N-player coalitions have a value, primarily because there has to be specific combinations of players using specific combinations of resources to any create value.  The mathematical function describing the value created by N-person coalitions is not super-additive because combinations of coalitions can increase or decrease psychological incentives/motives and hence values.  Each player can possibly gain an amount that is smaller than he/she could have gained on his/her own.  There cannot be any dominant strategy.  Contrary to Nash's theories, there cannot be any allocation of possible utility functions to any player that will make the player be in a position to threaten to be come an egoist unless other players adopt that player's utility function – this is primarily because: 1) there has to be a minimum level of cooperation among specific combinations of coalitions in order to create any value for the firm  (for the game to have any value), and 2) utility functions can change depending on level of incentive-effects, actual productivity and psychological motivation provided by such incentives. Sub-sets of each game are cooperative in the sense that at least some of the players (eg. employees; investors) have to cooperate for any value (ie. increase in the company's share price) to be created.

    Furthermore, in this game, the number of agents and players is finite; there is imperfect information.  There are incomplete contracts – the EBI and most incentive contarcts are typically incomplete contracts.  There is substantial information asymmetry among the players – about the value of the EBIs, the true value of the minimum required employee effort, the future values of the underlying stock prices, confidentiality of the EBI agreement, possibility of coalitions, etc..  There are penalties for misinformation and fraud, and insider trading; and there are specific rules pertaining to information exchange (eg. disclosure rules, Regulation-FD, etc.).

    The game is an NTU game.  The firm typically cannot transfer its interests in the negotiation, or the EBIs or incentive programs.  The employees also typically cannot transfer their interests in the negotiation process or the EBIs, at this stage.

    In this game, the Core is empty primarily because there can be many coalitions that can block any payoff or improve any payoff.  Different combinations of executives and shareholders can collude to change the terms of EBIs and compensation plans, subject to existing securities laws, corporation laws and labor laws.  The government can create new laws that change the payoffs and tax consequences of EBIs and other incentive programs.  In some jurisdictions, shareholder approval is not required in order to change the terms of EBIs and or incentive plans.

Assume that:
"Negotiating-Employee" refers to employee that is negotiating the EBIs/ESOs; "Other-Employees" refers to other employees in the company.

**$F_e$** = dollar magnitude of penalties that may be allocated to the employee for insider trading and or improper disclosure, or other associated misconduct;
**$F_c$** = dollar magnitude of penalties that may be allocated to the company for insider trading and or improper disclosure, or other associated misconduct;
**$Q_i$** = the negotiation strategy of the *i*th player, which is a function of various component strategies denoted as $a_i$ and include the following: $a_1$ = lower or increase the ESO/EBI strike price; $a_2$ = shorten the vesting period; $a_3$ = include anti-dilution features; $a_4$ = include registration rights; $a_5$ = minimize tax consequences of EBI/ESO award and ESO/EBI exercise; $a_6$ = where possible, amend financial reporting to increase reported earnings and cash flow, and to maximize perceived value of the firm's equity; $a_7$ = minimize probability of detection; $a_8$ = minimize B, $a_9$ = change the minimum required effort level, $a_{10}$ = change basis for disclosure, ......... $a_n$).
**B** = minimum performance benchmarks that must be achieved before EBIs/ESOs are valuable.
$€_a$ = work effort contributed by employee.
**K** = strike price of the EBI.
$I_{oe}$ = information value of un-detectable collusion between the Negotiating-Employee and Other- Employees. Collusion includes sharing information about the firm's prospects, agreeing on minimum amounts/terms of incentive compensation, agreeing on capital budgeting decisions, etc..
**Φ** = dollar amount of losses that arise due to moral hazard that is attributable to the issuance of EBIs.

The negotiating employee's strategy will be to maximize value of ESO/EBI by pursuing strategy:

Max $Q(a_i...a_n) + S - K + I_{oe}$
Min $(€_a + F_e + B)$

The Negotiating-Employee will almost always be able to collude with Other-Employees to achieve desired results. Shareholders (S), cannot participate directly in the game but are represented by company management. However, shareholders can intervene in the game by exercising their corporate powers (special voting, seeking board seats, shareholder derivative lawsuits, etc.) – these shareholder actions will be referred to as $S_p$.

Differences between the company's management and shareholders result in moral hazard (**Φ**), which is calculated in dollars. Hence the shareholders' multi-objective strategy is as follows:

Min $C(X)^{(\pi * \lambda * \psi)}$;
Max $C(B+€_a)^{(\pi * \lambda * \psi)}$;

Where: $X = [Q(a_1, a_2, a_3, a_4, a_5, .....a_n) + F_c + Φ]$.

C(.) is a transformation function and reflects agency and the tempering effect of management's representation of shareholders; and the voting power of shareholders. Hence C(.) is a function of several factors including $S_p$, the percentage of equity owned by management, the percentage of equity owned by institutional investors, perceived efficiency of the company's corporate governance practices, etc.

In each game, there is the issue of demand and supply of EBIs; and the equilibrium amount of EBIs. On the demand side, not all employees are granted EBIs. The firm has to designate classes of employees that will receive EBIs and the managers then select employees that will be granted crtain amounts of EBIs. Most employees typically prefer to be granted EBIs because of the asymmetric payoff and potential windfall gains. Hence, demand is constrained by firm selectivity, manager selectivity, and applicable laws/statutes. The employee's problem

Stage-2.
**T**he second stage of the 'game' (post-negotiation) consists of several repeated games, each of which occurs during the fiscal/reporting quarter (each game lasts for three months, and begins n the first day of the quarterly reporting period). These second-stage games involve numerous players including the stock market, investors, securities regulators, external auditors, venture capital firms, internal auditors, management, executives, and employees. This game occurs because: a) the company discloses or is required to disclose information to investors and regulators, and b) there are inherent moral hazard and principal-agent problems in the use of ESOs/EBIs, c) employees seek the best opportunities to exercise their ESOs/EBIs. Each of these games is characterized as follows; d) there is substantial information asymmetry – pertaining to the value of information, the firm's prospects and opportunity set, and the value of the EBIs. Note that the primary objective of this game is dual and is the optimal exercise of the EBI and the minimization of associated tax consequences – which affects any related actions such as risk tolerance of executives, the firm's capital budgeting decisions, dividend policy, human resources policies, etc..

Each game is a non-cooperative game; and a non-zero-sum game. The rules of the game are established by disclosure laws (eg. SEC rules, FASB/IASB rules, Sarbanes-Oxley Act), stock exchange rules/laws, internal audit reviews, and corporate governance standards. Payoffs are largely determined by payoffs and outcome of Phase One. The value of the game in time *t* is affected by value of the game and coalitions in time *t*-1. The transaction costs and monitoring costs incurred by the players (ie. regulators, stock exchanges, and employees) can be substantial. Coalitions are possible and occur depending on the reporting period, duration of the EBIs/ESOs, vesting periods, minimum performance benchmarks, etc. The amount of benefits to all players can change and is a function of their efforts, time, etc.

Furthermore, in each game, the number of agents and players is finite; and there is imperfect information. There are incomplete contracts – the EBI and most incentive contracts are typically incomplete contracts. There is substantial information asymmetry among the players – about the value of the EBIs, the true value of the minimum required employee effort, the future values of the underlying stock prices, the true level of confidentiality of the EBI agreement, the possibility of coalitions, etc.. There are penalties for misinformation and fraud, and insider trading; and there are specific rules pertaining to information exchange (eg. disclosure rules, Regulation-FD, etc.).

Each game in each fiscal quarter is a TU game. The issuing company can transfer its economic interests in the EBIs or incentive programs through the use of derivative instruments and contracts. Each employee can also transfer his/her interests in, and utility from the EBIs at this stage primarily through the use of derivatives.

In each game, the Core is empty, and there can be many coalitions that can block any payoff or improve any given payoff, because of certain problems: a) the ability of regulators and the company to observe players' actions, b) truthfulness in disclosures by companies, c) employees perceptions of fairness of incentive compensation plans, d) actions of lobbying companies, e) changes in tax laws, f) the company's restrictions on employees' exercise of EBIs – this can be a major constraint, g) . Different combinations of executives, public investors and shareholders can collude to change the prices of the firm's shares or to increase the volatility of the shares. Different coalitions of employees can agree to exercise their EBIs at or around the same time – which can have significant information content, and thus will affect payoffs of all players. Different combinations of executives and shareholders can collude to change the terms of EBI contracts and incentive compensation plans, subject to existing securities laws, corporation laws and labor laws. The government can create new laws that change the payoffs and tax consequences of EBIs and other incentive programs. In some jurisdictions, shareholder approval is not required in order to change the terms of EBIs and or incentive plans.

In each game, there is the issue of demand and supply of EBIs; and the equilibrium amount of EBIs. Equilibrium cannot be properly defined in the traditional sense of matching 'demand' with "supply". In this instance, equilibrium in each game is a state that is defined by the exercise of that amount of EBIs (owned by one or more employees) that maximizes the firm's payoffs, the employee's payoff and the government's payoff. On the demand side, not all employees are granted EBIs; and employees that hold EBIs face an optimization problem.

Hence, demand is constrained by firm selectivity, manager selectivity, and applicable laws/statutes. Supply refers to the firm's restrictions on exercise of EBIs, the availability of capital with which the employee can exercise the EBI, and tax effect of EBIs on the fir, the company and the government.

The value of each game varies for each player and is a function of: a) the value of the EBI/ESO, plus or minus the value of Transaction Costs, Monitoring Costs, Tax Impact, etc., b) the value of volatility of the stock price that is created by information asymmetries and differences in knowledge, c) the utility/disutility derived from owning the ESO/EBI, or from being involved in hedging the ESO/EBI, or by issuing the ESO/EBI, or by being a market regulator. The solution to each N-payer game is a set of imputations that cover maximization or minimization of the value of the EBI, transaction costs, and monitoring costs.

The value of game to the company is:

$$G_c = [\mathbf{U}_c + \mathbb{C}_a - V_c - C_c - \Lambda_c - M_c - L_c - T_c]^{(\pi^* \psi^* \lambda)};$$

The Company's multi-objective function will be:
$\text{Min } \int_0^{T_c} \int_0^{H} \{V_c + T_c + M_c + L_c + C_c\}^{(\pi^* \lambda)} \partial \mathbf{H} \partial T_c$.
$\text{Min } (\mathbb{C}_r - \mathbb{C}_a)$
$\text{Max } [\mathbf{U}_c + \mathbb{C}_a - V_c - C_c - \Lambda_c - M_c - L_c - T_c]$

The value of the game to employee will be:

$$G_e = \text{Max} [\text{Max}(\{_0\int^{T_c}_0\int^H \{(U_c + \text{\textEuro}_a + V_c - T_c - \Lambda_c - M_c - L_c - C_c)^{(\pi^*\psi^*\lambda^*\Omega)}\} \partial H \partial T_c \}, 0), \{_0\int^{T_e}_0\int^H \{U_e + V_e - \text{\textEuro}_a - T_e - \Lambda_e - M_e - L_e - C_e\}^{(\pi^*\psi^*\lambda)} \partial H \partial T_e\}]$$

The following are the characteristics of these second-stage games, most of which contravene established Game Theory principles. Attempts at achieving the Nash Equilibria solution partly depends on the utility functions of the players, but there cannot be any Nash Equilibria. Two or more players cannot improve the Nash Equilibria solution simultaneously by varying their efforts, by increasing communication, by disclosing their strategies to other players. The Nash solution is not symmetric - because reversing the roles of the players will affect the solution. Attempts at achieving the Nash solution will likely be affected by the introduction/negotiation of irrelevant alternatives – such occurrences change players' motivations, incentive levels, strategies and ultimate objectives. Not all N-player coalitions have a value, primarily because there has to be specific combinations of players using specific combinations of resources to create value – eg. there must be at least one employee, and one employer, and one shareholder, and the firm must have some minimum amount of capital. The mathematical function that describes the value created by N-person coalitions is not super-additive because different combinations of coalitions can increase or decrease psychological incentives/motives and hence values. In each game, each player in a coalition can possibly gain an amount that is smaller than he/she could have gained on his/her own – depending on the value-creating potential of the company and the game, and the feasible/actual combinations of players and resources, and the applicability of rules. There cannot be any dominant strategy – primarily because the combination of players, applicability of rules and resources is constantly changing, and each player derives varying utility from each possible combination of players/rules/resources, and there is an infinite number of possible strategies. Contrary to game theory principles, there cannot be any allocation of possible utility functions to any player that will make the player be in a position to threaten to be come an egoist unless other players adopt that player's utility function – this is primarily because: 1) there has to be a minimum level of cooperation among specific combinations of coalitions in order to create any value (for the game to have any value), 2) utility functions change depending on the level/amount of incentives and psychological motivation, 3) the large number of possible and actual players reduces the bargaining power of any player, 4) many players are not constrained to participation in the games, and have the option to quit – investors can sell their shares, employees can forfeit their stock options, companies can cancel their incentive plans, the government can change regulations, etc., 5) no game is exactly the same as any prior game – conditions and the combinations of players/resources/capital vary in each period.

The number of agents and players in the games are relatively infinite and constantly changing – because the game includes capital markets participants, shareholders, bond-investors, employees, and various types of government regulators. The Stage-Two game is iterative. Knowledge gained by players typically increases over time with the number of games (players learn more about how to cheat, create value, shirk, and extract value).

Players' moves are not entirely independent of each other – players have access to information about other players' prior moves – in terms of monetization or hedging of EBIs, and or re-negotiation of EBIs. Each player's payoff partly depends on other player's moves – eg. each exercise of an EBI/ESO typically results in equity dilution and perhaps increased/decreased volatility of the share price, which affects the payoffs of other employees and players, and also affects the value of the firm. Each game involves substantial imperfect information, and considerable information asymmetry. Each game involves incomplete contracts – the EBIs are essentially incomplete contracts. There are penalties for misinformation and fraud, and insider trading. There are upper and lower bounds on players' actions and possible moves. These limitations arise from applicable laws (securities laws, corporations laws, trading rules, etc.), the company's restrictions on the exercise or monetization of EBIs, the company's capital, the employee's knowledge, the employee's perception of risk, the company's size and the nature of its operations, distance and time. There are specific rules pertaining to information exchange – securities laws, Regulation FD, disclosure laws, etc..

4. ESOs/EBIs, Production Functions And Service Functions.
ESOs/EBIs substantially change production-functions and service-functions of corporate entities primarily because ESOs/EBIs have substantial incentive effects and have positive/negative interactions with other factors. The ESO/EBI can be considered a factor of production/service because it can create increased motivation, Social Capital, Reputation Capital and Human capital, all of which are relevant elements/factors of production/service. Hanoch & Rothschild (1972); Agrell, Lindroth & Norrman (2000); Balasubramanian & Bhardwaj (2004); Brockett,

Cooper, Golden, Rousseau & Wang (2005); Kaplan & Henderson (2005); Swieringa (1998); Mosca & Viscolani (2004); Tollington (1998); Gilson & Schizer (2003); Jaffee & Freeman (2002); Katz (1990); Canterbery & Marvasti (2001); Camerer & Hogarth (1999); Miller (2001); Brown & Medoff (1978); Wilson & Jadlow (1982); Conrad, Sales, et al (2002); Beckmann (1977); Chirinko (2008); Tassey (2005); Neill (2003); Fedderke & Luiz (2008).

EBIs/ESOs are a valid factor of production or services because:
- ESOs/EBIs / create an 'Expectations Value' that affects executives, employees, suppliers and investors – the Expectations Value is akin to Social Capital, Reputation Capital and Human Capital – investors, supplier and customers rely on this Expectations Value as a measure of commitment and motivation. Current/prospective employees also rely on Expectations Value for future profits and motivation.
- ESOs/EBIs create psychological motive which in and of itself contributes positively or negatively to firm output – this element constitutes value (Psychological Value) that is distinct from the traditional Labor, Capital and Psychological-Value factors of production – the psychological incentive effect of ESOs/EBIs is a new, substantial and identifiable element of production/service in most companies;
- ESOs/EBIs constitute an Information Asset (distinct from traditional notions of Labor and Capital) because their values largely depend on assessments of information about the firm's future prospects;
- ESOs'/EBIs' risk-sharing function constitutes a factor of production/service because it determines and or substantially influences how much risk the firm and employees are willing to take, the risk profile of the company, the ability of the firm to accurately deploy capital and labor; and
- Most ESOs/EBIs have vesting requirements, and there are substantial switching costs when employees separate from the company – thus, the vesting requirement by itself constitutes a factor of production that is distinct from the traditional labor and capital inputs, because vesting implies performance, and the vesting terms determine the value of the vesting input.
- Most employee incentive plans (eg. stock option plans and employee stock ownership plans) are functionally similar to collective bargaining agreements, which in turn have been shown to affect worker productivity and factors of production.

Venture capital companies play an active role in entrepreneurship: a) they provide functional capital and risk capital – cash for expansion and development, b) they provide Social Capital – they make introductions ands referrals, and facilitate transactions, and establish relationships that otherwise would be difficult or near impossible for small companies to develop, c) they provide Reputational Capital - with the investment and backing of a VC firm, suppliers, bankers, and customers are more likely to take the startup company more seriously and do business with it, Admati & Pfleiderer (1994); d) they provide a risk reduction mechanism that has value – by reducing the perceived risk of companies, providing guidance and additional financing; e) Exit mechanism – in many cases VCs also provide an exit mechanism, because they typically have strong relationships with securities brokerage firms that underwrite IPOs, and also have good industry contacts that facilitate sale to another company (See: Raith (2003); Gilson & Schizer (2003). Hence, VC firms essentially function like employees of startups and significantly extend the boundaries of the startup firm, and change the production functions or service functions of firms. American VC firms invest almost exclusively in various forms of Convertible Preferred Stock (see: Gilson & Schizer (2002) and Bratton (2002)). However in other countries, VC firms are more likely to use other securities/instruments – some of these differences can be explained by adverse selection – as explained by Cumming & McIntosh (2002). Hence, the CPS held by VC firms is akin to a bundle consisting of a debt instrument plus a series of ESOs/EBIs. This is because the Convertible Preferred Stock ("CPS") often have anti-dilution features and "death spiral" features, such that if the startup company does not perform well, the VC firm gets control of the company (but without the bankruptcy process that arises in the case of debt instruments). Unfortunately, the literature on EBIs/ESOs has not addressed the impact of such "synthetic EBIs/ESOs" on firm performance and employee motivation. Perceived excessive dilution by CPS de-motivates employees, and is likely to make entrepreneur-manager to take more risk. Furthermore, CPS increases information asymmetry because the CPS terms typically require mandatory periodic disclosure to the VC firm, VC participation on the board of directors, and on occasion, the VCs ability to veto certain expenses or transactions; whereas employees that own EBIs/ESOs (and contribute capital – intellectual capital, etc.) don't get such privileges.

The traditional production or service function has the form:

$y = f(x_1, x_2, .., x_m)$, which can be modified to account for joint production. EBIs/ESOs (and some types of incentives) contravene some of the fundamental assumptions implicit in the theory of production/service functions. The following are the major assumptions of production./service functions, and why they are contravened by EBIs/ESOs.

*Assumption #1:* **the production/service factors ($X_1, X_2, X_3, ....X_n$) are defined to be non-negative.**
This assumption is not valid because EBIs/ESOs can cause de-motivation and can have negative effects on output. The law of variable proportions does not hold as usual because of the positive or negative incentive effects of ESOs/EBIs and the potentially unlimited ESO/EBI profits. EBIs/ESOs change the returns-to-scale from associated factors of production/service primarily because of their incentive and tax effects.

*Assumption #2* **if $y = f(X_1, X_2, .., X_m)$, then $\partial y/\partial x_i = f_i \geq 0$, and $\partial^2 y/\partial x_i^2 = f_{ii} < 0$.**
For these conditions to hold, the units of the production/service factor must be 'continuous' (infinitely divisible), and if not, the marginal product of any factor cannot be mathematically expressed in the form of derivatives. Some factor units are not infinitely divisible, and as a factor of production/service, the positive/negative incentive effects of EBIs/ESOs are not always 'continuous', primarily because: a) there are vesting requirements and exercise restrictions, b) there are 'jumps' in the stock price. Thus, traditional production/service theory is not valid. The above-mentioned assumption of production/service functions erroneously implies diminishing marginal productivity of the *i*th factor, but EBIs/ESOs can have, and often do have the exact opposite effect (increasing marginal productivity) in companies and industries that are growing, ie. where: 1) the firm is growing, is profitable and producing positive operating cash flow and is earning above its cost of capital, 2) in most instances, the EBIs/ESOs returns are unlimited. Thus,
if $y = f(X_1, X_2, .., X_m)$, then, $\partial y/\partial x_i = f_i \geq 0$, and $\partial^2 y/\partial x_i^2 = f_{ii} \cong 0$

The law of diminishing returns does not hold as usual because of the incentive effects of EBIs/ESOs and potentially unlimited EBIs/ESOs profits. As more units are produced (or more units of services are performed), the firm's operating cash flows can increase directly with volume, and employees are motivated to perform better by the positive co-movements of volume, cash flow and equity prices. ESOs/EBIs also influence whether marginal productivity of other production factors is decreasing or rising.

*Assumption #3:* **the production/service function $f(x)$ is finite, non-negative, real-valued and single-valued for all non-negative and finite x.**
However, even if ESOs/EBIs (as a factor), is non-negative and finite: 1) $f(x)$ can be infinite because of the incentive effects and open-ended structure of most existing EBIs/ESOs, 2) $f(x)$ can have more than one value if the relationship is non-linear, 3) $f(x)$ can be negative depending on the nature of the interaction between ESOs/EBIs and the capital and labor factors.

*Assumption #4:* **the production function $f(x_1, x_2, x_3, ....x_n) = 0$, when $x_1, x_2, x_3, ....x_n$ are all equal to zero.**
This implies that if there are no inputs, then there will not be any output. This assumption is not valid in some situations, such as outsourcing (or distribution), where the contractor (or distributor) has certain pre-specified incentives. Furthermore, ESOs/EBIs and incentives can have negative values, which when combined with other factors, produce positive or negative values, and under such conditions, the following relationships exist:

$f(x_1, x_2, x_3, ....x_n) \leq 0$, where $x_1, x_2, x_3, ....x_n$ are all equal to zero; and
$f(x_1, x_2, x_3, ....x_n) \geq 0$, where $x_1, x_2, x_3, ....x_n$ are all less than zero.

*Assumption #5:* **if factor $X \geq X_1$, then $f(X) \geq f(X_1)$.**
That is, an increase in inputs does not decrease output. ESOs/EBIs (and some types of incentives) contravene this monotonicity condition. There are situations where ESOs (and other incentives) can have detrimental effects, reduce motivation, and create negative behaviors. Conditions under which increasing ESO/EBI grants will reduce output include the following: 1) where the ESOs/EBIs are deep out-of-the-money and there is an industry downturn or firm-specific problems, 2) in a labor intensive production environment, issuance of more ESOs/EBIs will execerbate employee-management friction, 3) issuing ESOs/EBIs in lieu of cash payments in a unionized company.

*Assumption #6:* **the production/service function $f(\ )$ is continuous and twice-continuously differentiable everywhere in the interior of the production set.**

This condition is not valid because ESOs/EBIs (as a factor of production/service) can be discrete.

*Assumption #7:* **If the production/service function is defined by: $y = f(x_1, x_2, x_3, \ldots x_n)$;
then the set $V(y) = \{x \mid f(x) \geq y\}$ is a convex set.**

This assumption is not valid because in their present form, ESOs/EBIs can generate potentially unlimited returns/profits.

*Assumption #8:* **if the production/service function is defined by $y = f(x_1, x_2, x_3, \ldots x_n)$;
then the set $V(y)$ is closed and non-empty for any $y > 0$.**

This assumption is not valid because in their present form, ESOs/EBIs have unlimited and potentially infinite returns.

5. Conclusion.

In their present form, most ESOs/EBIs (including indexed ESOs) are not an efficient method of compensating or motivating employees. The ESO/EBI can be considered a factor of production/service because it can create increased motivation, Social Capital, Reputation Capital and Human capital. ESOs/EBIs change production functions and service functions drastically. While Game-Theory is flawed (and is more limited than was previously thought), it has been used in this article to analyze some of the economic and psychological effects of interactions among various parties in ESO/EBI transactions. The negotiation, grant, holding and exercise of ESOs/EBIs involves a two-stage game; however, these processes contravene most existing and defined principles of Game Theory.